\documentclass[conference]{IEEEtran}
\usepackage[utf8]{inputenc}

\usepackage[cmex10]{amsmath}
\usepackage{amsthm}
\usepackage{cite}
\usepackage{array}
\usepackage{multirow}
\usepackage{booktabs}
\usepackage{url}
\usepackage{algorithm2e}

\usepackage{float}
\usepackage{tikz}
\ifcsname basic.pgf\endcsname
  \expandafter 
\fi
\expandafter\gdef\csname basic.pgf\endcsname{loaded}

\usetikzlibrary{circuits.logic.US}
\usetikzlibrary{shapes.geometric}
\tikzset{
  multiplexer/.style={
    draw,
    trapezium,
    shape border uses incircle,
    shape border rotate=270,
    minimum size=14pt
  }
}
\tikzstyle{branch}=[fill,shape=circle,minimum size=3pt,inner sep=0pt]

\SetAlgorithmName{Alg.}{}{} 
\SetAlgoCaptionSeparator{.}

\newcommand{\rfig}[1]{Fig.\,\ref{#1}}
\newcommand{\rtab}[1]{Tab.\,\ref{#1}}
\newcommand{\ralg}[1]{Alg.\,\ref{#1}}

\newtheorem{definition}{Definition}

\newcommand\xor\oplus
\renewcommand\bar\overline
\makeatletter
\newsavebox{\@euflag}
\sbox{\@euflag}{\raisebox{-9mm}{\resizebox{!}{1.1cm}{\begin{tikzpicture}
\fill[fill={rgb,255:red,0; green,51; blue,153}] (-27, -18) rectangle (27,18);  
\pgfmathsetmacro\inr{tan(36)/cos(18)}
\foreach \i in {0,1,...,11} {
  \begin{scope}[shift={(30*\i:12)}]
    \fill[fill={rgb,255:red,255;green,204;blue,0}] (90:2)
	\foreach \x in {0,1,...,4} { -- (90+72*\x:2) -- (126+72*\x:\inr) };
  \end{scope}
}
\end{tikzpicture}
}}}
\newcommand\euflag{\usebox{\@euflag}\xspace}
\makeatother

\begin{document}

\title{
  Generic and Universal Parallel Matrix Summation with a Flexible Compression Goal for Xilinx FPGAs
}

\author{
  \IEEEauthorblockN{
    Thomas B. Preu{\ss}er$^\dagger$
  }
  \IEEEauthorblockA{
    Department of Computer Science\\
    Technische Universit\"{a}t Dresden, Germany\\[.2ex]
    thomas.preusser@utexas.edu
  }
}

\maketitle

{\renewcommand{\thefootnote}{$\dagger$}
  \footnotetext{%
    \euflag\parbox[t]{.752\linewidth}{%
      This work has been revised under the funding of the
      Marie Skłodowska-Curie Grant Agreement No.\,751339 
      of the Eu\-ropean Union's Framework Programme
      for Research and Innovation Horizon 2020 (2014-2020)
      at Xilinx Ireland.
    }
  }
}

\begin{abstract}
Bit matrix compression is a highly relevant operation in computer
arithmetic. Essentially being a multi-operand addition, it is the
key operation behind fast multiplication and many higher-level
operations such as multiply-accumulate, the computation of the dot
product or the implementation of FIR filters. Compressor
implementations have been constantly evolving for greater efficiency
both in general and in the context of concrete applications or
specific implementation technologies. This paper is building on this
history and describes a generic implementation of a bit
matrix compressor for Xilinx FPGAs, which does not require a generator
tool. It contributes FPGA-oriented metrics for the evaluation of
elementary parallel bit counters, a systematic analysis and partial
decomposition of previously proposed counters and a fully implemented
construction heuristic with a flexible compression target matching the
device capabilities. The generic implementation is agnostic of the
aspect ratio of the input matrix and can be used for multiplication
the same way as it can be for single-column population count
operations.
\end{abstract}


\IEEEpeerreviewmaketitle

\section{Introduction}\label{secIntroduction}
Bit matrix compression is a key operation in efficient computer
arithmetic. It was already the general concept behind the fast
multiplier suggested by Wallace \cite{wallace:1964} and the multiplier
schemes discussed by Dadda \cite{dadda:1965} as well as the carry shower
circuits that were proposed by Foster and Stockton \cite{foster:1971} and
further discussed by Swartzlander
\cite{swartzlander:1973}. Besides multiplication and population
counting, matrix compression can naturally subsume fused operations
such as multiply-accumulate \cite{kwon:2002} or the computation of
complete dot products. The dot product by itself is a key operation in
applications such as digital filters \cite{mirzaei:2006} or neural
networks \cite{umuroglu:2017}. The latter work specifically advocates
binarized neural networks for their efficient implementation on
FPGAs. In this approach, the dot product summation degenerates into
a high-fanin population count, which was the motivation to evaluate
this work in the context of matrices with extreme aspect ratios.

\begin{figure}
  \begin{minipage}[b]{.6\linewidth}
    \begin{center}\begin{tikzpicture}[scale=.45,line width=1pt]
  \foreach \x in {-1,0,1,2,4,5,6,7}
    \filldraw[color=gray](\x,4) circle(.26);
  \node at (3,4) {$\times$};
  \draw (-1.5,3.5) -- (7.5,3.5);
  \foreach \y in {0,1,2,3}
    \foreach \x in {0,1,2,3}
      \filldraw(\x+\y+1,\y) circle(.26);
  \node at (-1,0) {$+$};
  \draw (-1.5,-.5) -- (7.5,-.5);
  \foreach \x in {0,1,...,7}
    \filldraw(\x,-1) circle(.26);
  \draw (-1.5,-1.5) -- (7.5,-1.5);
  \draw (-1.5,-1.7) -- (7.5,-1.7);
\end{tikzpicture}\\(a)\,Multiplication\end{center}
  \end{minipage}\hfill
  \begin{minipage}[b]{.3\linewidth}
    \begin{center}  \begin{tikzpicture}[scale=.45,line width=1pt]
    \foreach \x in {0,1,2,3,4,5,6}
      \filldraw(2,\x) circle(.26);
    \node at (0,0) {$+$};
    \draw (-.5,-.5) -- (2.5,-.5);
    \foreach \x in {0,1,2}
      \filldraw(\x,-1) circle(.26);
    \draw (-.5,-1.5) -- (2.5,-1.5);
    \draw (-.5,-1.7) -- (2.5,-1.7);
  \end{tikzpicture}\\(b)\,Popcount\end{center}
  \end{minipage}
  \caption{Shapes of Matrices to be Compressed for Implementations of
    (a)\,Multiplication and (b)\,Population Count}\label{figMatrixShapes}
\end{figure}
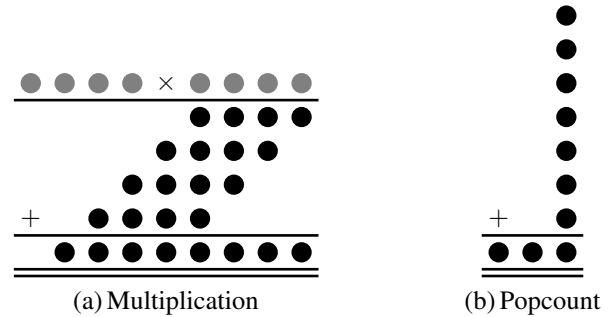

Already the early, well established use cases exemplify the
ubiquity of bit matrix
reduction and the diversity of shapes of the input matrices. On the
one hand, multiplication processes a matrix of partial bit products,
which assumes a skewed shape due to the increasing numerical weight of
the multiplier bits. The population count operation behind the carry
shower circuits, on the other hand, has to process a single column of
equally weighted bits. Both of these use cases are illustrated in
\rfig{figMatrixShapes}.

Ultimately, both of these use cases desire to compute the arithmetic
sum of all properly weighted input bits as a binary number. This goal
is achieved through two distinguishable steps: (1) the matrix
compression down to a height of two rows by making a massively
parallel use of elementary bit counters, and (2) the carry-propagate
addition to obtain the conventional binary result. At the interface
between these steps, the result is available in what is called a
carry-save representation, i.e., its total is distributed over two
addends. Higher-level operations fusing multiple compression steps
would typically try to copy this representation directly to the input
matrix of the next compression so as to avoid the carry-propagate
addition for all but the conclusive computation step. This approach
warrants a significant speedup as the logarithmic latencies of
efficient implementations of both the matrix compression and the final
carry-propagate addition are on the same order of magnitude.

\begin{figure}
  \centerline{\begin{tikzpicture}[scale=.45,line width=1pt]
  \foreach \x in {0,1,2,3,4,5,6,7,11} {
    \foreach \y in {-1,0,1,2}
      \filldraw(\x,\y) circle(.26);
    \filldraw(\x-1,-2) circle(.26);
    \draw[black,rounded corners](\x-.4,-0.4) rectangle(\x+.4,2.4);
    \draw [ultra thick] (\x,-1) -- (\x-1,-2);
  }
  \draw (-1.5,-0.6) -- (7.5,-0.6);
  \draw (9,-3) -- (9,3);
  \draw (9.5,-0.6) -- (11.5,-0.6);
  \draw (13,-1.5) rectangle (15, .5);
  \draw (14,-.5) node {FA};
  \draw[->] (11,2) -- (14.5,2) -- (14.5,.5);
  \draw[->] (11,1) -- (12,1) -- (12,1.5) -- (14,1.5) -- (14,.5);
  \draw[->] (11,0) -- (12.5,0) -- (12.5,1) -- (13.5,1) -- (13.5,.5);
  \draw (13.5,-1.5) -- (13.5,-2.5) -- (10,-2.5) -- (10,-2);
  \draw (14.5,-1.5) -- (14.5,-2) -- (11,-2) -- (11,-1);
\end{tikzpicture}}
  \caption{Carry-Save-Addition}\label{figCSA}
\end{figure}
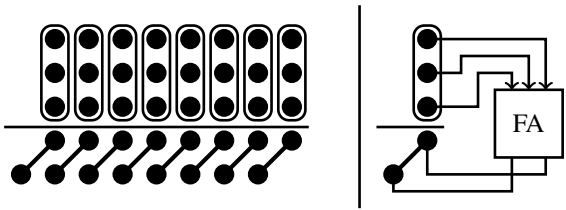
The matrix compression is implemented by reduction elements typically
referred to as parallel counters. The most basic of these counters is
the full adder reducing three inputs of the same weight to a two-digit
binary number:
\[x_0 + x_1 + x_2 = 2\cdot y_1 + y_0\quad\mbox{with }x_*, y_*\in\left\{0,1\right\}\]
where $y_1$ corresponds to the carry and $y_0$ to the sum
output. Applying full adders in parallel to
three matrix rows as shown in \rfig{figCSA} reduces them to
two rows with the same additive value. This 3-to-2 compression,
aka. carry-save addition, is performed within a single full-adder
delay independent of the width of the rows. More sophisticated
counters can give rise to higher compression ratios such as the
4-to-2-adders thoroughly reviewed by Kornerup \cite{kornerup:2002} or
the 5-to-2 compressor described by Kwon et
al. \cite{kwon:2002}. Besides full adders, they rely on instances of
what is called a generalized parallel counter whose input bits may
already have different numerical weights. In particular, their
compressor also uses $\left(2,3:1,1,1\right]$ elements where the notation of
$\left(p_{m-1}, \dots, p_0:q_{n-1}, \dots, q_0\right]$ defines the right-aligned
numbers of input and output bits, which the counter processes while
maintaining this invariant over the total sum:

\[\sum_{i=0}^{m-1}\left(2^i\cdot\sum_{j=0}^{p_i-1}x_{i,j}\right)=\sum_{i=0}^{n-1}\left(2^i\cdot\sum_{j=0}^{q_i-1}y_{i,j}\right)\]

The rather stiff notion of reducing the number of complete
rows is less suitable for irregular matrix shapes. In fact, it was
already broken up by Dadda \cite{dadda:1965} who applied full and half
adders for the reduction of a multiplication matrix exactly to the bit
columns where this was needed to reach the targeted row count of
the current reduction step. This flexible, goal-oriented placement of
counters can still be considered the state of the art. Heuristics and ILP
solvers have been used to optimize such compression solutions for
various input matrix shapes \cite{parandeh:2011,kumm:2014}.

In the remainder of this paper, we first give an overview of the
related work before establishing criteria for the counter evaluation
and defining a suitable and systematically completed set of parallel
bit counters for our FPGA
implementation. We then discuss the conclusive carry-propagate
addition and its integration with the preceding matrix compression to
define an efficient greedy construction of a matrix summation
implementation. Finally, we evaluate the generic synthesizable VHDL
implementation of our approach targeting a concrete Xilinx Zynq device
using Vivado.

\section{Related Work}
The scheduling of counters to build a compressor depends
naturally on the selection of available modules. It is the backing
technology that defines which counters can be implemented most
efficiently. A discussion of the choices for ASICs was composed by
Verma and Ienne \cite{verma:2007}. FPGA-targeted counters have been
most prominently proposed by Parandeh-Afshar et
al. \cite{afshar:2008,afshar:2009,parandeh:2011} as well as Kumm and
Zipf \cite{kumm:2014a,kumm:2014}. As this paper focuses on the
construction of compressors within a modern Xilinx FPGA fabric, it
will heavily build on the work of these latter two groups.

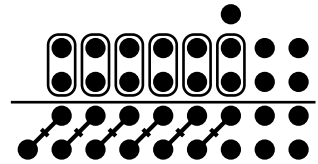
\begin{figure}
  \centerline{\begin{tikzpicture}[scale=.45,line width=1pt]
  \foreach \x in {0,1,2,3,4,5} {
    \foreach \y in {-1,0,1}
      \filldraw(\x,\y) circle(.26);
    \filldraw(\x-1,-2) circle(.26);
    \draw[black,rounded corners](\x-.4,-0.4) rectangle(\x+.4,1.4);
    \draw [ultra thick] (\x,-1) -- (\x-1,-2);
    \draw [ultra thick] (\x-0.6,-1.4) -- (\x-0.4,-1.6);
  }
  \foreach \y in {-2,2}
    \filldraw(5,\y) circle(.26);
  \foreach \x in {6,7} {
    \foreach \y in {-2,-1,0,1}
      \filldraw(\x,\y) circle(.26);
  }
  \draw (-1.5,-0.6) -- (7.5,-0.6);
\end{tikzpicture}}
  \caption{Compulsory Parallel Matrix Re-Shaping Using Half Adders}
  \label{figHAmust}
\end{figure}
A heuristic for constructing compressors for Altera devices was
proposed by Parandeh-Afshar et al. in 2008 \cite{afshar:2008}. They
used a single-pass heuristic selecting the most efficient from a selection
of parallel counter that would fit into the work still to do by
the compression step starting from the least-significant and
proceeding to the most-significant bit position. The compression goal
was a matrix of, at most, three rows. This relaxed goal definition
exploits the fact that ternary adders map well onto modern FPGA
architectures. It also has the tremendous benefit that half adders can
be avoided altogether. Half-adders only have a reshaping function and
do not reduce the number of bits in the matrix. As shown in
\rfig{figHAmust}, they must be used to reshape an almost done two-row
matrix in parallel so that it can accommodate just one more carry
efficiently. This pressure disappears with a goal of three rows.

In their follow-up work \cite{afshar:2009}, Parandeh-Afshar et
al. start considering mapping counters to the broader structural
context of an Altera Adaptive Logic Module (ALM) rather than assuming
an indifferent pool of lookup tables (LUTs). This enables them to
exploit the carry-chain links between adjacent LUT stages for fast and
yet more capable counters. Finally \cite{parandeh:2011}, they tie
individual counters together by merging the carry output of one module
with the carry input of another into one LUT stage. While this,
indeed, reduces LUT usage, it also creates unwieldy structures that
severely limit the mobility of individual counters during the logic
placement, which complicates the routing optimization to be performed
by the tools. Last but not least, this work also looks into a
generalization for Xilinx architectures.

Kumm and Zipf \cite{kumm:2014a}, on the other hand, proposed counter
designs that are a natural fit for the slices of four LUTs that are
found in Xilinx architectures. Like Parandeh-Afshar et al., they aim
at a three-row compression goal. While they provide convincing delay
estimates for using their proposed counters, they do not use it for a
sharp cut in the selection of implementation modules. Besides an
extended counter selection, they also propose to consider 4:2 adders
in the construction of a compressor. In a subsequent work
\cite{kumm:2014}, they substitute the heuristic compressor
construction by an integer linear programming (ILP)
optimization. While they can demonstrate a consistent reduction of the
LUT usage and the number of compression stages, the ILP running time
remains prohibitive for all but desperate workflows.

Targeting Xilinx devices, the work by Kumm and Zipf is the natural
foundation we build on. We adopt their useful counters but also
decompose, classify and generalize them to construct a systematic
selection of modules, from which the implementation can pick the most
suitable and capable instances. While we also
use their \emph{efficiency} metric, we complement it with the
additional metrics of \emph{strength} and \emph{slack} so as to enable
a directed selection of the most beneficial counters in the
compressor construction. Performing an in-workflow construction at the
time of the RTL synthesis, we rely on a heuristic that is closely
related to the one by Parandeh-Afshar et al.

\section{Counter Evaluation}
For the evaluation of the building blocks of a compressor, the
generalized parallel counters, we derive several performance metrics
based on their physical properties. The goal of this evaluation is to
firstly define the selection of counters that is to be used in the
compressor construction and secondly to prioritize among them as long as
multiple choices are technically feasible.

As a first criterion, we will use the estimated counter delay for a
hard exclusion of candidates. So as to ensure roughly balanced bit
delays after each compression step, counters are not allowed to add
any extra signal paths over the general-purpose routing
network beyond what is needed to feed its inputs and to forward
its outputs. Counters are, however, allowed to grow beyond a single
LUT by using slice internal signal paths, in particular, the carry
chain. The delay on the carry chain links is negligible in comparison
to general-purpose routing paths. We only ensure that a counter
is constrained to a slice, which corresponds to a maximum of four
LUTs. While all the GPCs collected by Kumm and Zipf also fit into the
bounds of a slice \cite{kumm:2014}, quite a few of them feature
secondary carry signals over the general-purpose routing in parallel
to the carry-chain link. We exclude these counters explicitly.

In terms of physical dimension, the total number of counter inputs,
outputs and the occupied area in terms of LUTs are of interest. Given
the GPC $\left(p_{m-1}, \dots, p_0:q_{n-1}, \dots, q_0\right]$, we use:

\[\begin{array}{ll}
p = \sum_{i=0}^{m-1}p_i & \mbox{\quad-- the total number of inputs}\\
q = \sum_{i=0}^{n-1}q_i & \mbox{\quad-- the total number of outputs}\\
k                      & \mbox{\quad-- the number of occupied LUTs}
\end{array}\]

Performance metrics are derived from these physical characteristics.

\begin{definition}
  The \emph{efficiency} of the generalized parallel counter
  $\left(p_{m-1}, \dots, p_0:q_{n-1}, \dots, q_0\right]$ is the
  quotient of its achieved reduction of the number of bit signals and
  the number of LUTs it occupies:
  \[E = \frac{p-q}{k}.\]
\end{definition}
This notion of efficiency was previously used by Kumm and Zipf
\cite{kumm:2014}. It reflects signal reduction achieved in relation to
the hardware investment. Giving preference to more efficient counters
will optimize the constructed result with respect to silicon area.

\begin{definition}
  The \emph{strength} of the generalized parallel counter
  $\left(p_{m-1}, \dots, p_0:q_{n-1}, \dots, q_0\right]$ is the ratio
  of its input bit count vs. its output bit count:
  \[S = \frac{p}{q}.\]
\end{definition}
The strength metric captures the asymptotic height reduction of a
large bit matrix when exclusively using a specific counter in a single
compression step. Giving preference to stronger counters emphasizes a
small number of compression steps as a construction goal.

\begin{definition}
  The \emph{(arithmetic) slack} of the generalized parallel counter
  $\left(p_{m-1}, \dots, p_0:q_{n-1}, \dots, q_0\right]$ captures the
  share of the numeric range representable by the output bits that
  cannot be used:
  \[A = 1-\frac{1+\sum_{i=0}^{m-1} 2^i\cdot p_i}
               {1+\sum_{i=0}^{n-1} 2^i\cdot q_i}\]
\end{definition}
\begin{figure}
  \centerline{\begin{tikzpicture}[scale=.45,line width=1pt]
  \foreach \y in {0,1,3,4,5,6,7,8,9}
    \filldraw(2,\y) circle(.26);
  \foreach \y in {1,2,3,4,9}
    \filldraw(1,\y) circle(.26);
  \filldraw(0,0) circle(.26);
  \draw[black,rounded corners](2-.4,6.6) rectangle(2+.4,9.4);
  \draw[black,rounded corners](2-.4,4.6) rectangle(2+.4,6.4);
  \draw (-.5,4.5) -- (2.5,4.5);
  \draw [ultra thick] (2,4) -- (1,3);
  \draw [ultra thick] (2,3) -- (1,2);
  \draw [ultra thick] (2-0.6,2.6) -- (2-0.4,2.4);
  \draw[black,rounded corners](1-.4,1.6) rectangle(1+.4,4.4);
  \draw (-.5,1.5) -- (2.5,1.5);
  \draw [ultra thick] (1,1) -- (0,0);
\end{tikzpicture}}
  \caption{Phantom Carries by Slacky Counters}
  \label{figPhantomCarry}
\end{figure}
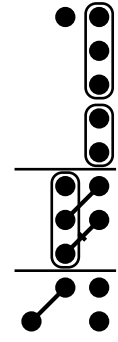
The slack captures the coding inefficiency of a counter's output. It
is counterproductive for reducing the number of bits in the
matrix. Accumulating slack within a compression network may even
result in phantom carries. Refer to the compression depicted by
\rfig{figPhantomCarry}. While the two-row result looks as if it might
produce a carry into the $2^3$ position, this is, in fact, not
possible because the maximum numerical value of the original input is
only $1\cdot2+5\cdot1=7$. This misconception suggested by the dot diagram is
created by the half adder (a $\left(2:1,1\right]$-counter), which
cannot produce a result value of three and hence leaves one out of
four possible outputs unused. Note that the slack of a functionally
correct counter is never negative as this would imply that there
are large input totals that cannot be recoded into the available
output bits\footnote{%
  Counters with a negative slack may render useful when (positive)
  slack has accumulated in the computation. For instance, a half adder
  that has no arithmetic chance to produce a carry degenerates into a
  plain XOR gate for the sum output. This gate can be viewed as a
  simple $\left(2:1\right]$-counter with an obvious negative
  slack. This pathway is not investigated any closer in this paper.
}. Giving preference to counters with no
or, at least, smaller slack minimizes the chance to introduce phantom
signals to the constructed compressor.

\begin{figure}
  \begin{center}\input{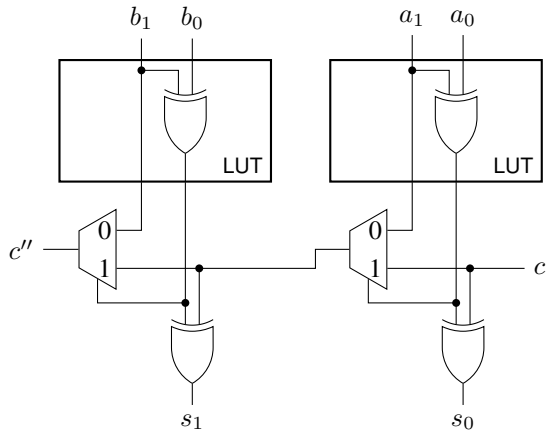}

\begin{tikzpicture}[circuit logic US,scale=.9]
\foreach \x in {0,1} {
  \coordinate (BASE) at (4-4*\x,0);
  \node[rectangle,draw,thick,minimum width=80pt,minimum height=46pt] (lut\x) at ($(BASE)+(0,3.4)$) {};
  \node[above left] at (lut\x.south east) {\sffamily\footnotesize LUT};
  \node[multiplexer, rotate=180] (muxcy\x) at ($(BASE)+(-1,1.5)$) {};
  \node at ([xshift=-5pt]muxcy\x.south west) {0};
  \node at ([xshift=-5pt]muxcy\x.north west) {1};
  \node[xor gate, rotate=-90] (xorcy\x) at ($(BASE)+(.4,0)$) {};
  \coordinate (lut{\x}o5) at ([xshift=-10pt]lut\x.south);
  \coordinate (lut{\x}o6) at (xorcy\x.input 2 |- lut\x.south);
  \draw (xorcy\x.input 2) -- (lut{\x}o6);
  \draw (muxcy\x.north) -- ($(muxcy\x.north)-(0,10pt)$) node (n1) {} -| (xorcy\x.input 2|-n1) node[branch] {};
  \draw (xorcy\x.input 1|-muxcy\x.north west) node[branch] {} -- (xorcy\x.input 1);
  \draw (muxcy\x.south west) -| (lut{\x}o5);
  \node (s\x) at ([yshift=-16pt]xorcy\x.output) {$s_\x$};
  \draw (xorcy\x.output) -- (s\x);
}
\node (c0) at ([xshift=64pt]muxcy0.north west) {$c$};
\draw (c0) -- (muxcy0.north west);
\draw (muxcy0.east) -- ++(left:.5) |- (muxcy1.north west);
\node (c2) at ([xshift=-24pt]muxcy1.east) {$c''$};
\draw (muxcy1.east) -- (c2);

\foreach \x in {0,1} {
  \node[xor gate, rotate=-90] (lut{\x}xor) at (lut{\x}o6|-lut0) {};
  \draw (lut{\x}xor.output) -- (lut{\x}o6);
}
\node (b0) at ($(lut{0}xor.input 1)+(0,32pt)$) {$a_0$};
\node (a0) at (lut{0}o5|-b0) {$a_1$};
\node (b1) at ($(lut{1}xor.input 1)+(0,32pt)$) {$b_0$};
\node (a1) at (lut{1}o5|-b1) {$b_1$};
\foreach \x in {0,1} {
  \draw (b\x) -- (lut{\x}xor.input 1);
  \draw (a\x) -- (lut{\x}o5);
  \draw (lut{\x}xor.input 2) -- ([yshift=10pt]lut{\x}xor.input 2) node (n1) {} -- (lut{\x}o5|-n1) node[branch] {};
}
\end{tikzpicture}\end{center}
  \caption{Atom $(2,2)$}
  \label{figAtom22}
\end{figure}
\begin{figure}
  \begin{center}\input{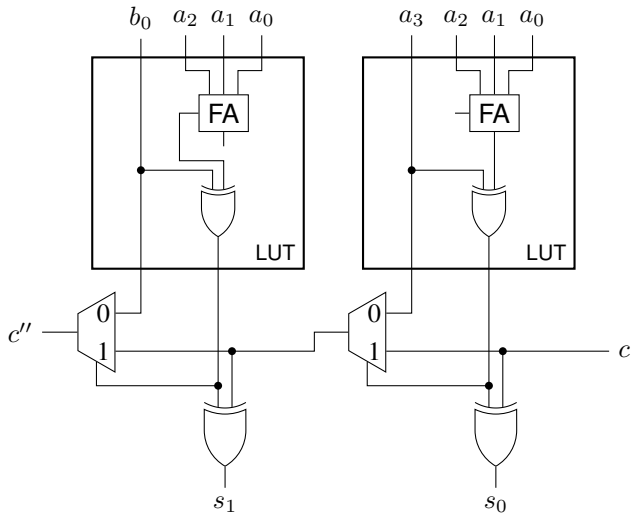}

\begin{tikzpicture}[circuit logic US,scale=.9]
\foreach \x in {0,1} {
  \coordinate (BASE) at (4-4*\x,0);
  \node[rectangle,draw,thick,minimum width=80pt,minimum height=80pt] (lut\x) at ($(BASE)+(0,4)$) {};
  \node[above left] at (lut\x.south east) {\sffamily\footnotesize LUT};
  \node[multiplexer, rotate=180] (muxcy\x) at ($(BASE)+(-1.5,1.5)$) {};
  \node at ([xshift=-5pt]muxcy\x.south west) {0};
  \node at ([xshift=-5pt]muxcy\x.north west) {1};
  \node[xor gate, rotate=-90] (xorcy\x) at ($(BASE)+(.4,0)$) {};
  \coordinate (lut{\x}o5) at ([xshift=-24pt]lut\x.south);
  \coordinate (lut{\x}o6) at (xorcy\x.input 2 |- lut\x.south);
  \draw (xorcy\x.input 2) -- (lut{\x}o6);
  \draw (muxcy\x.north) -- ($(muxcy\x.north)-(0,10pt)$) node (n1) {} -| (xorcy\x.input 2|-n1) node[branch] {};
  \draw (xorcy\x.input 1|-muxcy\x.north west) node[branch] {} -- (xorcy\x.input 1);
  \draw (muxcy\x.south west) -| (lut{\x}o5);
  \node (s\x) at ([yshift=-16pt]xorcy\x.output) {$s_\x$};
  \draw (xorcy\x.output) -- (s\x);
}
\node (c0) at ([xshift=100pt]muxcy0.north west) {$c$};
\draw (c0) -- (muxcy0.north west);
\draw (muxcy0.east) -- ++(left:.5) |- (muxcy1.north west);
\node (c2) at ([xshift=-24pt]muxcy1.east) {$c''$};
\draw (muxcy1.east) -- (c2);

\node (v{0}) at ($(lut{0}o5|-lut0.north)+(0,16pt)$) {$a_3$};
\node (v{1}) at (lut{1}o5|-v{0}) {$b_0$};
\foreach \x in {0,1} {
  \node[xor gate, rotate=-90,scale=.8] (lut{\x}xor) at ($(lut{\x}o6|-lut\x)+(0,-20pt)$) {};
  \draw (lut{\x}xor.output) -- (lut{\x}o6);
  \node[rectangle,draw,shape border uses incircle] (fa{\x}) at ($(lut{\x}xor.input 1)+(0,32pt)$) {\sffamily FA};
  \foreach \u in {0,1,2} {
    \node (u) at ($(fa{\x}|-v{0})+(-\u*16pt+16pt,0pt)$) {$a_\u$};
    \draw (u) -- ++(90:-22pt) -| ([xshift=-\u*6pt+6pt]fa{\x}.north);
  }
  \draw (v{\x}) -- (lut{\x}o5);
  \draw (lut{\x}xor.input 2) -- ++(0,8pt) node (n1) {} -- (lut{\x}o5|-n1) node[branch]{};
}
\draw (fa{0}.south) -- (lut{0}xor.input 1);
\draw (fa{0}.west) -- ++(left:6pt);
\draw (fa{1}.south) -- ++(90:-6pt);
\draw (fa{1}.west) -- ++(left:8pt) -- ++(90:-20pt) -| (lut{1}xor.input 1);
\end{tikzpicture}\end{center}
  \caption{Atom $(1,4)$}
  \label{figAtom14}
\end{figure}
\begin{figure}
  \begin{center}\input{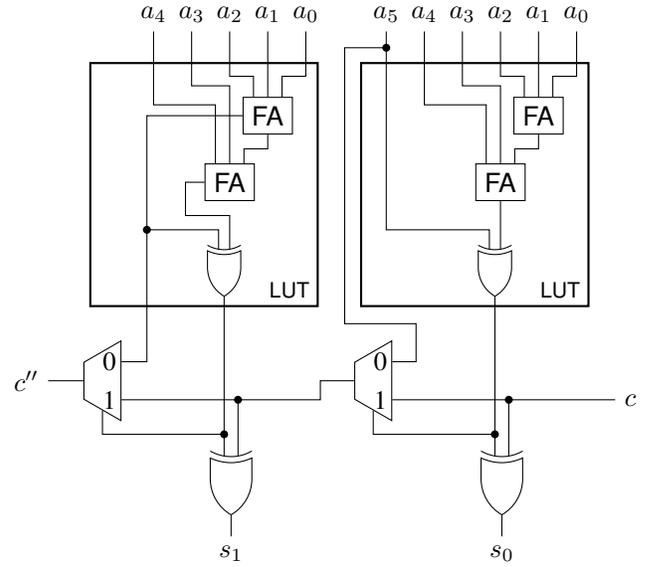}

\begin{tikzpicture}[circuit logic US,scale=.9]
\foreach \x in {0,1} {
  \coordinate (BASE) at (4-4*\x,0);
  \node[rectangle,draw,thick,minimum width=86pt,minimum height=92pt] (lut\x) at ($(BASE)+(0,4.4)$) {};
  \node[above left] at (lut\x.south east) {\sffamily\footnotesize LUT};
  \node[multiplexer, rotate=180] (muxcy\x) at ($(BASE)+(-1.5,1.5)$) {};
  \node at ([xshift=-5pt]muxcy\x.south west) {0};
  \node at ([xshift=-5pt]muxcy\x.north west) {1};
  \node[xor gate, rotate=-90] (xorcy\x) at ($(BASE)+(.4,0)$) {};
  \coordinate (lut{\x}o5) at ([xshift=-24pt]lut\x.south);
  \coordinate (lut{\x}o6) at (xorcy\x.input 2 |- lut\x.south);
  \draw (xorcy\x.input 2) -- (lut{\x}o6);
  \draw (muxcy\x.north) -- ($(muxcy\x.north)-(0,10pt)$) node (n1) {} -| (xorcy\x.input 2|-n1) node[branch] {};
  \draw (xorcy\x.input 1|-muxcy\x.north west) node[branch] {} -- (xorcy\x.input 1);
  \node (s\x) at ([yshift=-16pt]xorcy\x.output) {$s_\x$};
  \draw (xorcy\x.output) -- (s\x);
}
\draw (muxcy1.south west) -| (lut{1}o5);
\node (c0) at ([xshift=100pt]muxcy0.north west) {$c$};
\draw (c0) -- (muxcy0.north west);
\draw (muxcy0.east) -- ++(left:.5) |- (muxcy1.north west);
\node (c2) at ([xshift=-24pt]muxcy1.east) {$c''$};
\draw (muxcy1.east) -- (c2);

\coordinate (iref) at ($(lut0.north)+(0,20pt)$);
\foreach \x in {0,1} {
  \node[xor gate, rotate=-90,scale=.8] (lut{\x}xor) at ($(lut{\x}o6|-lut\x)+(0,-36pt)$) {};
  \draw (lut{\x}xor.output) -- (lut{\x}o6);
  \node[rectangle,draw,shape border uses incircle] (fa{\x}r) at ($(lut{\x}xor.input 1)+(0,28pt)$) {\sffamily FA};
  \node[rectangle,draw,shape border uses incircle] (fa{\x}l) at ($(lut{\x}xor.input 1)+(16pt,56pt)$) {\sffamily FA};
  \foreach \u in {0,1,2} {
    \node (u) at ($(fa{\x}l|-iref)+(-\u*16pt+16pt,0pt)$) {$a_\u$};
    \draw (u) -- ++(90:-26pt) -| ([xshift=-\u*6pt+6pt]fa{\x}l.north);
  }
  \draw (fa{\x}l.south) -- ++(90:-6pt) -| ([xshift=6pt]fa{\x}r.north);
  \node (u) at ($(fa{\x}r|-iref)+(-16pt,0pt)$) {$a_3$};
  \draw (u) -- ++(90:-30pt) -| (fa{\x}r.north);
  \node (u) at ($(fa{\x}r|-iref)+(-32pt,0pt)$) {$a_4$};
  \draw (u) -- ++(90:-38pt) -| ([xshift=-6pt]fa{\x}r.north);
}
\node (u0) at ($(fa{0}r|-iref)+(-48pt,0pt)$) {$a_5$};
\draw (fa{0}r.south) -- (lut{0}xor.input 1);
\draw (lut{0}xor.input 2) -- ++(90:8pt) -| (u0);
\draw (muxcy0.south west) -- ++(right:10pt) -- ++(90:16pt) -- ++(left:30pt) node (n1) {} -- ([yshift=6pt]lut0.north-|n1) node (n2) {} -- (u0|-n2) node[branch] {};

\draw (fa{1}r.west) -- ++(left:8pt) -- ++(90:-14pt) -| (lut{1}xor.input 1);
\draw (fa{1}l.west) -| (lut{1}o5);
\draw (lut{1}xor.input 2) -- ++(90:8pt) node (n1) {} -- (lut{1}o5|-n1) node[branch]{};

\end{tikzpicture}\end{center}
  \caption{Atom $(0,6)$}
  \label{figAtom06}
\end{figure}
Kumm and Zipf have proposed several counters that map perfectly into
the slice structure found in modern Xilinx devices since
generation~5. While not identified as such,
most of them are actually instances of a more general
concept that composes those 4-column counters from the 2-column atoms
shown from \rfig{figAtom22} through \rfig{figAtom06}. Any two of these atoms can
be combined arbitrarily into a slice to form nine different
counters. Both constituting atoms are exclusively connected through
the carry chain. The initial carry chain input at the lower
significant atom can be used to input an additional bit of weight
one. Note that this is physically not possible for atom
$(0,6)$ since the LUT bypass used by $a_5$ conflicts with driving an
external carry input. All of the counters constructed this way produce
a five-bit binary number as a single-row result.

\begin{table}
  \caption{Composable Whole-Slice Counters}
  \label{tabSliceCounters}
  \begin{minipage}{\linewidth}
    \renewcommand{\thefootnote}{\fnsymbol{footnote}}
    \renewcommand{\thempfootnote}{\fnsymbol{mpfootnote}}
    \begin{center}
    \begin{tabular}{cc@{~}c@{~}c@{\;}cc@{~}c@{~}c@{\;}cc@{~}c@{~}c@{\;}c}\toprule
      \multirow{2}{*}{\textbf{Atoms}} &
      \multicolumn{4}{c}{$(\dots,2,3)$} &
      \multicolumn{4}{c}{$(\dots,1,5)$} &
      \multicolumn{4}{c}{$(\dots,0,6)$} \\
                    & $E$ &$S$&$A$&& $E$ &$S$&$A$&& $E$ &$S$& $A$& \\\midrule
      $(2,2,\dots)$ & 1   &1.8&0&\footnotemark[2] & 1.25&2  &0& & 1.25&2  &$\frac{1}{32}$& \\
      $(1,4,\dots)$ & 1.25&2  &0& & 1.5 &2.2&0&\footnotemark[3] & 1.5 &2.2&$\frac{1}{32}$&\footnotemark[3] \\
      $(0,6,\dots)$ & 1.5 &2.2&0& & 1.75&2.4&0& & 1.75&2.4&$\frac{1}{32}$&\footnotemark[3] \\\bottomrule
    \end{tabular}
    \end{center}
    \footnotetext[2]{Standard 4-bit RCA.}
    \footnotetext[3]{Counters proposed by Kumm and Zipf \cite{kumm:2014a,kumm:2014}.}
  \end{minipage}
\end{table}

The performance metrics of these composable whole-slice counters are
summarized by \rtab{tabSliceCounters}. On the left top, the
combination of two $(2,2)$-atoms with an additional carry input
essentially yields a four-bit ripple-carry adder (RCA). It reduces the
number of active bits by one for each invested LUT, hence its
efficiency of $1$. While an arbitrarily wide RCA would have a strength
of $2$, the limitation of the accepted carry paths to a slice limits
it to $1.8$. Both efficiency and strength grow systematically as one
of the weight-2 inputs is replaced by two weight-1 inputs to yield the
$(1,4)$- and the $(0,6)$-atoms. They complete the set of reasonable 2-column
atoms, which are allowed to contribute a maximum numeric value of 6 in
addition to the input on the carry chain. The remaining
alternative of a $(3,0)$-atom is reasonably subsumed by a single-LUT
implementation of a full adder.

Note that the advantage of the $(0,6)$-atom is impacted in the
low-significant position where a structural
resource hazard within the slice prevents the functional utilization
of an additional carry input. Only being able to feed a constant of zero, there
are no improvements in terms of efficiency and strength over the
$(1,4)$-atom in this position. Rather arithmetic slack is
introduced as the output code space cannot be utilized completely. For
the illustration of the strength metric, observe that the exclusive use
of $(0,6,1,5)$- and $(0,6,0,6)$-counters on a large bit matrix will
deflate twelve input rows to five output rows, hence $S=\frac{12}{5}$.

There is one more whole-slice counter that is adopted from Kumm and
Zipf \cite{kumm:2014} which cannot be decomposed into the atoms
extracted above: the $\left(1,3,2,5:1,1,1,1,1\right]$-counter. We discard
some other of their proposals:
\begin{itemize}
\item
  all counters such as the $\left(2,0,4,5:1,1,1,1,1\right]$-counter
  that impose an additional routing delay by driving carry-like
  signals over the general-purpose routing rather than the carry
  chain, and
\item
  the $\left(1,5:1,1,1\right]$- and $\left(6:1,1,1\right]$-counters
  that they have mapped to the
  carry chain tying them to the lower significant half of
  the slice without being able to utilize the higher significant part.
\end{itemize}

\begin{figure}
  \centerline{\input{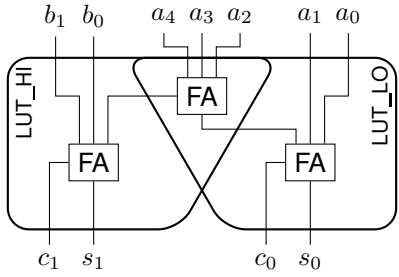}

\begin{tikzpicture}[scale=.9]
\node[trapezium,draw,thick,rounded corners=8pt,trapezium left angle=90,trapezium right angle=120, minimum width=36mm] (hi) at (-1.6,0) {};
\node[below left,rotate=90] at (hi.north west) {\sffamily\footnotesize LUT\_HI};
\node[trapezium,draw,thick,rounded corners=8pt,trapezium left angle=120,trapezium right angle=90, minimum width=36mm] (lo) at ( 1.6,0) {};
\node[above left,rotate=90] at (lo.north east) {\sffamily\footnotesize LUT\_LO};

\node[rectangle,draw,below] (fa_hi) at (hi) {\sffamily FA};
\node[rectangle,draw,below] (fa_lo) at (lo) {\sffamily FA};
\node[rectangle,draw] (fa_mid) at (0,20pt) {\sffamily FA};

\node[above] (a3) at ($(fa_mid|-hi.north)+(0,10pt)$) {$a_3$};
\node (a2) at ([xshift= 16pt]a3) {$a_2$};
\node (a4) at ([xshift=-16pt]a3) {$a_4$};
\node (a1) at (fa_lo|-a3) {$a_1$};
\node (a0) at ([xshift= 16pt]a1) {$a_0$};
\node (b0) at (fa_hi|-a3) {$b_0$};
\node (b1) at ([xshift=-16pt]b0) {$b_1$};

\draw (a0) -- (a0|-fa_mid) -| ([xshift=6pt]fa_lo.north);
\draw (a1) -- (fa_lo.north);

\draw (a2) -- ++(90:-13pt) -| ([xshift=6pt]fa_mid.north);
\draw (a3) -- (fa_mid.north);
\draw (a4) -- ++(90:-13pt) -| ([xshift=-6pt]fa_mid.north);

\draw (b0) -- (fa_hi.north);
\draw (b1) -- (b1|-fa_mid) -| ([xshift=-6pt]fa_hi.north);

\draw (fa_mid.west) -| ([xshift=6pt]fa_hi.north);
\draw (fa_mid.south) -- ++(90:-6pt) -| ([xshift=-6pt]fa_lo.north);

\coordinate (oref) at ($(lo.south)-(0,6pt)$);
\draw (fa_lo.south) -- (fa_lo|-oref) node[below] {$s_0$};
\draw (fa_lo.west) -- ++(left:8pt) node (n1) {} -- (n1|-oref) node[below] {$c_0$};
\draw (fa_hi.south) -- (fa_hi|-oref) node[below] {$s_1$};
\draw (fa_hi.west) -- ++(left:8pt) node (n1) {} -- (n1|-oref) node[below] {$c_1$};

\end{tikzpicture}}
  \caption{Implementation of the $\left(2,5:1,2,1\right]$-Counter}
  \label{figCounter25}
\end{figure}
\begin{table}
  \caption{Evaluation of Floating Counters}
  \label{tabFloatingCounters}
  \begin{center}
    \begin{tabular}{rc@{~}c@{~}c}\toprule
      \textbf{Counter} & $E$ & $S$ & $A$ \\\midrule
      $\left(3:1,1\right]$     & 1  & 1.5 & 0 \\
      $\left(6:1,1,1\right]$   & 1  &  2  & $\frac{1}{8}$\\
      $\left(2,5:1,2,1\right]$ & 1.5& 1.75& 0 \\\bottomrule
    \end{tabular}
  \end{center}
\end{table}
While complex, slice-based counters promise high values in strength
and efficiency, they are typically too bulky for achieving a covering
of the bit matrix that is as exhaustive as possible. For this purpose,
more flexible, smaller LUT-based counters are needed. The most
elementary among these is the full adder, i.e. a
$\left(3:1,1\right]$-counter. Using both of its outputs, this can be
implemented within a single LUT. The $\left(6:1,1,1\right]$-counter is
adopted from Brunie et al. \cite{brunie:2013}. It occupies three
LUTs and is very effective for reducing the height of a
singular peek column. Finally, we propose the novel $\left(2,5:1,2,1\right]$-counter depicted in \rfig{figCounter25}. It utilizes only two LUTs to do the
work of three full adders within a single logic level. As shown, the
sum and the carry computations of one of these full adders are
distributed among the two LUTs and merged with the exclusively local full
adders. Implementing two 5-input functions within every involved LUT
utilizes them fully. The performance figures of all of these
\emph{floating} counters whose LUTs can be placed freely are listed in
\rtab{tabFloatingCounters}. Note that the
$\left(2,5:1,2,1\right]$-counter competes very favorably in this group.

\section{Final Carry-Propagate Addition}
The ultimate result of the parallel compression by counters must
finally undergo a carry-propagate addition to obtain the total as a
conventional binary number. The traditional two-row compression goal
implied by Wallace \cite{wallace:1964} and Dadda \cite{dadda:1965} can
be relaxed to three rows if ternary adders are well supported on the
targeted hardware. This was exploited by Parandeh-Afshar et
al. \cite{afshar:2008} as well as Kumm and Zipf \cite{kumm:2014a}. We
will go one step further and define an even more flexible compression
goal.

\begin{figure}
  \centerline{\input{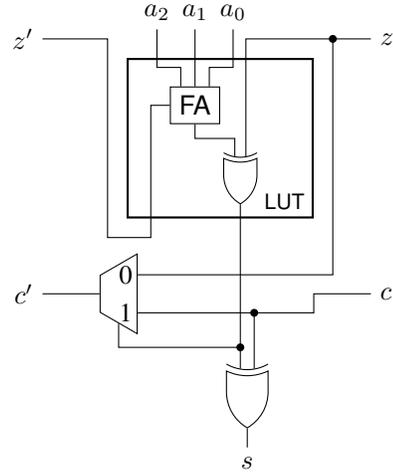}

\begin{tikzpicture}[circuit logic US,scale=.9]

\node[rectangle,draw,thick,minimum width=70pt,minimum height=60pt] (lut) at (0,3.8) {};
\node[above left] at (lut.south east) {\sffamily\footnotesize LUT};
\node[multiplexer, rotate=180] (muxcy) at (-1.5,1.5) {};
\node at ([xshift=-5pt]muxcy.south west) {0};
\node at ([xshift=-5pt]muxcy.north west) {1};
\node[xor gate, rotate=-90] (xorcy) at (.4,0) {};
\draw(muxcy.east) -- ([xshift=-24pt]muxcy.east) node (c1) {}; \node[left] at (c1) {$c'$};
\draw(muxcy.south west) -| ($(lut.north east)+(8pt,8pt)$) node[branch] (z0) {};
\draw(muxcy.north west) -- (lut.east|-muxcy.north west) |- ([xshift=24pt]lut.east|-muxcy.east) node (c0) {}; \node[right] at (c0) {$c$};
\draw(muxcy.north) -- ++(90:-10pt) node (n1) {} -| (xorcy.input 2|-n1) node[branch] {};
\draw(xorcy.input 1|-muxcy.north west) node[branch] {} -- (xorcy.input 1);
\draw(xorcy.output) -- ++(90:-8pt) node[below] {$s$};

\node[rectangle,draw,below left] (fa) at ([yshift=-12pt]lut.north) {\sffamily FA};
\node[xor gate, rotate=-90,scale=.8,left] (xor) at ([yshift=6pt]xorcy.input 2|-lut.south) {};
\draw(fa.west) -- ++(left:8pt) |- ($(lut.south west)-(8pt,8pt)$) |- (c1|-z0) node[left] {$z'$};
\draw(fa.south) -- ++(90:-6pt) -| (xor.input 2);
\draw(xor.output) -- (xorcy.input 2);
\draw(xor.input 1) |- (c0|-z0) node[right] {$z$};

\coordinate(iref) at ([yshift=12pt]fa.north|-lut.north);
\draw(fa.north)--(iref) node[above] (a1)  {$a_1$};
\draw([xshift= 6pt]fa.north)--++(90:8pt)-|([xshift= 16pt]iref) node[above] {$a_0$};
\draw([xshift=-6pt]fa.north)--++(90:8pt)-|([xshift=-16pt]iref) node[above] {$a_2$};

\end{tikzpicture}}
  \caption{Element of a Ternary Adder in the Xilinx Carry Chain}
  \label{figTernary}
\end{figure}
The ternary adder implementation on a Xilinx carry chain requires a
secondary carry that cannot utilize a direct fast carry-chain link. As
shown in \rfig{figTernary}, the computation of this secondary carry
$z'$  at a bit position does not depend on the incoming secondary carry
$z$. Thus, no lengthy and slow combinational path is created but
rather only one extra intermediate routing delay is implied in
total. This additional delay is roughly equivalent to
introducing a separate compression stage by parallel full adders that
would achieve a reduction from three down to two rows as
well. However, the ternary adder offers an increased functional density.

Instead of targeting a fixed compression goal as the input to the
carry-propagation stage, we allow a flexible goal that is inferred
in the process of the right-to-left greedy heuristic for counter
placement. This process will not further consider the placement of
counters in a column if its effective height, i.e. its remaining input
bits plus the outputs from counters previously placed in this
compression step, can be processed by the carry-propagate stage. The
acceptable bit height depends on the previous history of column
heights. In the very first, least-significant column, a height of four
bits is acceptable as the input $z$ can be re-purposed as a fourth
input. Unfortunately, $z$ and $c$ compete for the same link to the
general-purpose routing network so that a fifth input fails on a
structural hazard.

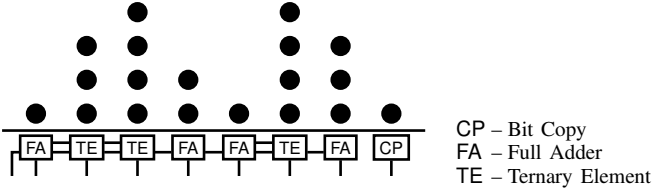
\begin{figure}\footnotesize
  \begin{tikzpicture}[scale=.45,line width=1pt]
  \newcommand\s{1.5}
 
  \node[rectangle,draw,scale=.8] (e0) at (\s*7,-1) {\sffamily CP};
  \node[rectangle,draw,scale=.8] (e1) at (\s*6,-1) {\sffamily FA};
  \node[rectangle,draw,scale=.8] (e2) at (\s*5,-1) {\sffamily TE};
  \node[rectangle,draw,scale=.8] (e3) at (\s*4,-1) {\sffamily FA};
  \node[rectangle,draw,scale=.8] (e4) at (\s*3,-1) {\sffamily FA};
  \node[rectangle,draw,scale=.8] (e5) at (\s*2,-1) {\sffamily TE};
  \node[rectangle,draw,scale=.8] (e6) at (\s*1,-1) {\sffamily TE};
  \node[rectangle,draw,scale=.8] (e7) at (\s*0,-1) {\sffamily FA};

  \draw([yshift=-4pt]e1.west)--([yshift=-4pt]e2.east);
  \draw([yshift= 4pt]e2.west)--([yshift= 4pt]e3.east);
  \draw([yshift=-4pt]e2.west)--([yshift=-4pt]e3.east);
  \draw([yshift=-4pt]e3.west)--([yshift=-4pt]e4.east);
  \draw([yshift=-4pt]e4.west)--([yshift=-4pt]e5.east);
  \draw([yshift= 4pt]e5.west)--([yshift= 4pt]e6.east);
  \draw([yshift=-4pt]e5.west)--([yshift=-4pt]e6.east);
  \draw([yshift= 4pt]e6.west)--([yshift= 4pt]e7.east);
  \draw([yshift=-4pt]e6.west)--([yshift=-4pt]e7.east);
  \draw([yshift=-4pt]e7.west)--++(left:6pt)--++(90:-20pt) node (oref) {};
  
  \foreach \x in {0,...,6,7} {
      \filldraw(\s*\x,0) circle(.26);
      \draw(e\x.south) -- (e\x.south|-oref);
  }
  \foreach \x in {1,2,3,5,6}
      \filldraw(\s*\x,1) circle(.26);
  \foreach \x in {1,2,5,6}
      \filldraw(\s*\x,2) circle(.26);
  \foreach \x in {2,5}
      \filldraw(\s*\x,3) circle(.26);

  \draw (\s*0-1,-.5) -- (\s*7+1,-.5);
  
\end{tikzpicture}\hfill
  \begin{tabular}[b]{l@{\;--\;}l}
  \sffamily CP & Bit Copy \\
  \sffamily FA & Full Adder \\
  \sffamily TE & Ternary Element
  \end{tabular}
  \caption{Example of an Acceptable Ragged Carry-Propagate Input}
  \label{figRagged}
\end{figure}
Behind a ternary adder element, the next column will be considered
done if it is no higher than three bits. Counters are placed
aggressively if the height target has not yet been reached. This way,
the resulting column height may drop below the acceptable height. This
allows benefits to be passed on to the next column. For instance,
assume the rightmost column to have a height of 5, and a
$\left(2,5:1,2,1\right]$ counter is scheduled for its compression. In
the next compression stage, no more counter will be placed into this
column. Due to the fact that it does not even leave a carry for the
final addition, the weight-2 column now only needs to be compressed to a
height of 4. Similarly, full adders that do not produce a secondary
carry will be used if the number of remaining bits permits. See
\rfig{figRagged} for an example of a ragged carry-propagate input that
is acceptable by the proposed flexible approach.

\begin{table}
  \caption{Construction of the Carry-Propagate Stage}
  \label{tabCPConstruction}
  \small\sffamily
  \begin{tabular}{ccccccc}\toprule
    \textbf{Carries $\backslash$ Height } &$0$&$1$&$2$&$3$&$4$&$>4$\\\midrule
    $0$& $\emptyset$ & CP & FA & FA & TE & n/a \\
    $1$&     CP      & FA & FA & TE & TE & n/a \\
    $2$&     FA      & FA & TE & TE & n/a& n/a \\\bottomrule
  \end{tabular}
\end{table}
\begin{algorithm}\footnotesize
  \KwData{height[WIDTH-1 \dots 0] -- effective column heights}
  anchor  := 0\;
  carries := 0\;
  \SetKwProg{myproc}{Procedure}{}{}
  \myproc{update\_anchor{}}{
    \While{(anchor $<$ WIDTH) and (height[anchor] $\le$ 4)\\\quad and (height[anchor] + carries $\le$ 5)}{
      schedule carry-propagate module (\rtab{tabCPConstruction})\;
      carries := (carries + height[anchor])/2\;
      anchor  := anchor + 1\;
    }
  }

  update\_anchor()\;
  \While{anchor $<$ WIDTH}{
    // another compression stage, try in order of preference\\
    \For{counter in COUNTERS}{
      \For{pos := anchor \dots W-counter.width}{
        \While{counter fits at pos}{
          schedule  counter\;
        }
      }
    }
    schedule pipeline registers as requested\;
    update\_anchor()\;
  }
 \caption{Greedy Compressor and Summation Construction}
 \label{algAnchor}
\end{algorithm}
\rtab{tabCPConstruction} and \ralg{algAnchor} illustrate the
interaction between the counter scheduling and the flexible
carry-propagate stage. Carry-propagate modules are selected on the
basis of the effective column height and the carries received through
the carry propagation itself as shown by \rtab{tabCPConstruction}. If
the effective column height does not fit any of the available
elements, the parallel compression by counters has to continue and
counters are to be placed starting at the determined \texttt{anchor}
position. Pipeline registers may be scheduled to buffer the
active bit signals after any compression stage. The corresponding
request must currently be posed by the designer who specifies the
stage count, which is expected to meet timing within the targeted
constraints.

\section{Experimental Evaluation}
The described algorithm was implemented directly in synthesizable VHDL
code. The available counters are characterized by signature records
stored in an array that is sorted with respect to the preferred
performance metric for the purpose of counter scheduling. The concrete
schedule comprising the counter placements and the composition of the
conclusive carry-propagate addition is computed by a designated
function, which takes an array of the initial column heights as its
input. Its output is the representation of the schedule in a flat
\texttt{integer\_vector}, which directs the actual module
instantiations within a \texttt{for-generate} loop.

We have used the described algorithm together with the proposed
selection of counters and the suggested final adder construction to
implement several matrix summations. The analyzed use cases range from
high, single-column population counts over high, dual-column inputs to
a $16\times16$-bit multiplication matrix for reference. A multiplier
would typically not be implemented within the general LUT-based FPGA
fabric but would rather utilize the hardwired $25\times18$-bit
multipliers of the DSP48E blocks.

All measurements were obtained using Vivado 2016.4 targeting an
XC7Z045-FFG900-2 device. This device is found on the ZC706 evaluation
board, which was also recently used by Umuroglu et
al. \cite{umuroglu:2017}. Delay measurements are taken from a pure
combinational implementation without added pipeline stages in a
register sandwich that determines the timing constraint. Timing
targets are defined in a process of nesting intervals until a failed
and an accomplished timing goal are reached that are no more than
0.1\,ns apart. The last met timing is reported. Area results for the
summation are extracted from the hierarchical utilization report. For
each use case, three schemes of the counter selection for the
construction of the compression stage have been evaluated: precedence
with respect to (a) efficiency, (b) strength, and (c) their
product. The arithmetic slack was used as last decision criterion only.

\begin{figure}
  \centerline{\includegraphics[width=\linewidth]{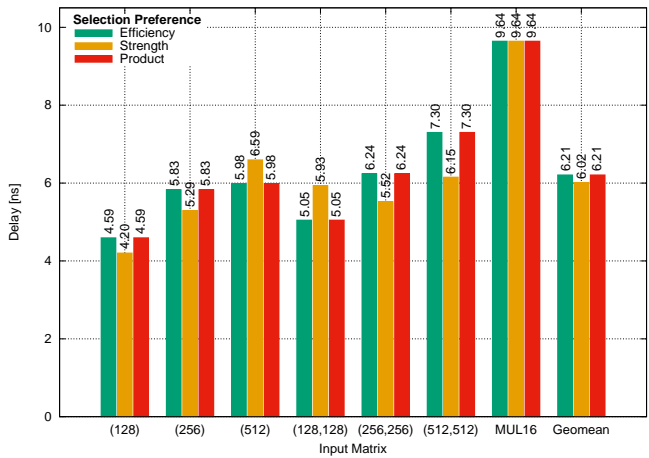}}
  \caption{Total Combinational Summation Delays}
  \label{figDelay}
\end{figure}
\rfig{figDelay} depicts the determined combinational delays of the
different matrix summations. The general tendency that higher or wider
input matrices demand more time for computation is obvious. However,
note that the strength-driven construction produces an unexpectedly
slow solution for the input matrix of two 128-bit columns. Its
schedule, indeed, differs significantly as it is dominated by
$\left(6:1,1,1\right]$- rather than the
$\left(2,5:1,2,1\right]$-counter more frequently employed by the other
approaches. Nonetheless, in the overall comparison the strength metric
tends to produce the fastest solutions.

\begin{table}
  \begin{center}
    \begin{tabular}{lc@{\,}c@{\,}c@{\,}ccc@{\,}c@{\,}c@{\,}cc}\toprule
      \multirow{2}{*}{\textbf{Matrix}}
      &\multicolumn{5}{c}{Efficiency / Product}&\multicolumn{5}{c}{Strength}\\
      &FA & $(2,5)$ & $(6)$ & Slice & Stages
      &FA & $(2,5)$ & $(6)$ & Slice & Stages\\\midrule
      (128)	&	4&3&22&5&4&	2&0&25&5&3\\
      (256)	&	6&1&49&12&4&	7&0&49&12&4\\
      (512)	&	7&5&97&26&5&	6&1&101&26&5\\
      (128,128)	&	4&29&18&14&5&	4&2&46&13&4\\
      (256,256)	&	8&58&37&29&6&	3&0&98&28&5\\
      (512,512)	&	11&116&78&59&7&	4&0&197&59&6\\
      MUL16	&	15&2&1&27&3&	12&1&2&28&3\\\bottomrule
    \end{tabular}
  \end{center}
  \caption{Scheduled Counters and Compression Stages}
  \label{tabCounters}
\end{table}
The soft $16\times16$-multiplier matrix appears to produce an
extraordinarily long delay with all solution approaches, especially
when realizing that it also only has a total of 256 input bits. It
suffers from its wide result, which is computed within a final 32-bit
carry-propagate adder. As can be seen in \rtab{tabCounters}, all
computed solutions are also
heavily dominated by whole-slice counters with only a few interspersed
floating ones. This suggests that the delay implications even of short
monolithic carry chains should be investigated more thoroughly in the
future. Note that the schedules optimized for strength and for the
efficiency-strength product were, in fact, identical in all the
presented use cases.

\begin{figure}
  \centerline{\includegraphics[width=\linewidth]{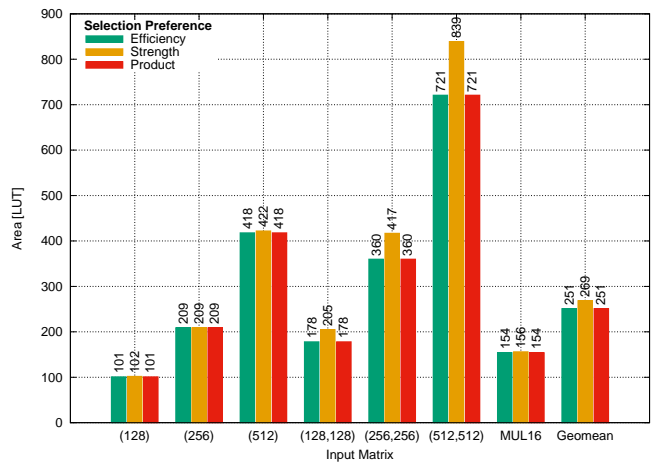}}
  \caption{Total Summation Area}
  \label{figArea}
\end{figure}
The area consumption of the summation solutions is shown in
\rfig{figArea}. It clearly shows the tradeoffs imposed by the counter
selection on different use cases in comparison to the delay
figures. While the summation of the multiplication matrix is the
slowest, it also has the most compact solution among the cases with
256 input bits. Here, the greater efficiency of the whole-slice
counters takes effect. Also note that the strength-driven selection
has to pay an area premium for achieving a certain speed gain over the
other approaches.

\begin{figure}
  \centerline{\includegraphics[width=\linewidth]{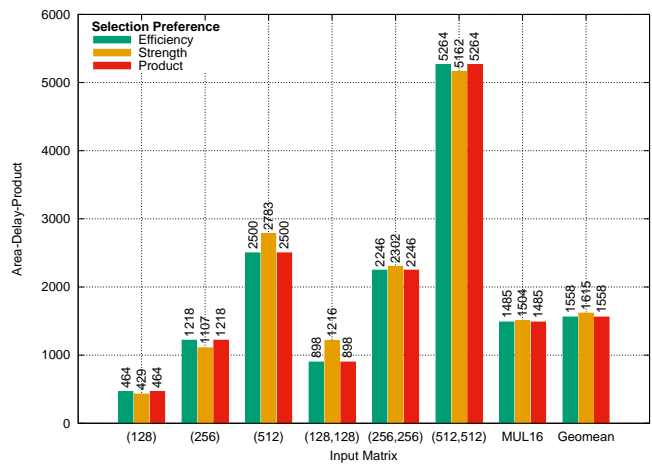}}
  \caption{Area-Delay-Product of Matrix Summations}
  \label{figAreaDelay}
\end{figure}
Using the combined area-delay product shown in \rfig{figAreaDelay} as
the quality metric, the differences between the selection approaches
diminish further. There is no clear winner, and a preference towards
area efficiency or speed should be selected explicitly so that
the most appropriate solution can be constructed.

For a concrete practical reference, recollect the population count
synthesized by Umuroglu et al. for FINN \cite{umuroglu:2017}. The
population count is one of the key operations in their binarized
neural network application. Their HLS implementation of a 128-bit
instance, however, is pipelined so as to meet the 200\,MHz clock
target and occupies a total of 376~LUTs. All compressors generated by
our very feasible and efficient greedy approach achieve this goal on
the same device in a single cycle of a 200\,MHz clock with only
slightly more than a quarter of the resources. This demonstrates that
providing such a critical operation possibly as a builtin function of
the HLS compiler is worth more than a consideration.

\newpage
\section{Conclusions}\label{secConclusions}
This paper has described a VHDL-implemented generic matrix summation
module that can be used universally for the implementation of
operations as different as population counting, dot product
computation or integer multiplication. The implementation is backed by
a set of parallel counters that has been derived and extended from
previous works by Parandeh-Afshar et al. and Kumm et al. The proposed
approach further features a novel flexible interface between the
parallel matrix compression and the conclusive carry-propagate
addition. The complete approach has been implemented specifically
targeting modern Xilinx devices. It has been shown that the
underlying runtime-efficient greedy construction of the matrix
summation is a valuable opportunity for an operation to be provided as
a builtin function of the high-level synthesis.

\newpage
\bibliographystyle{IEEEtran}
\bibliography{paper}

\end{document}